\documentclass[longauth]{aaemma} 

\usepackage[english]{babel}
\usepackage{graphicx,color}
\usepackage{color}
\usepackage[usenames,dvipsnames,svgnames,table]{xcolor}
\usepackage{natbib}
\bibpunct{(}{)}{;}{a}{}{,} 

\usepackage[small]{caption}

\usepackage{amssymb}
\usepackage{amsmath}
\usepackage[left,modulo]{lineno}
\usepackage{multirow, soul}

\begin{document}
%

\newcommand{\arcdeg}{$^{\circ}$}
\newcommand{\arcs}{$^{\prime\prime}$}
\newcommand{\arcm}{$^\prime$}
\newcommand{\g}{$\gamma$}
\renewcommand{\d}{$^\circ$}

\title{Probing the gamma-ray emission from HESS J1834$-$087 using H.E.S.S. and {\textit Fermi} LAT observations}

\subtitle{}

\author{H.E.S.S. Collaboration
\and A.~Abramowski \inst{1}
\and F.~Aharonian \inst{2,3,4}
\and F.~Ait Benkhali \inst{2}
\and A.G.~Akhperjanian \inst{5,4}
\and E.~Ang\"uner \inst{6}
\and G.~Anton \inst{7}
\and M.~Backes \inst{8}
\and S.~Balenderan \inst{9}
\and A.~Balzer \inst{10,11}
% ...
}

\institute{
Universit\"at Hamburg, Institut f\"ur Experimentalphysik, Luruper Chaussee 149, D 22761 Hamburg, Germany \and
Max-Planck-Institut f\"ur Kernphysik, P.O. Box 103980, D 69029 Heidelberg, Germany \and
Dublin Institute for Advanced Studies, 31 Fitzwilliam Place, Dublin 2, Ireland \and
National Academy of Sciences of the Republic of Armenia, Yerevan  \and
Yerevan Physics Institute, 2 Alikhanian Brothers St., 375036 Yerevan, Armenia \and
Institut f\"ur Physik, Humboldt-Universit\"at zu Berlin, Newtonstr. 15, D 12489 Berlin, Germany \and
Universit\"at Erlangen-N\"urnberg, Physikalisches Institut, Erwin-Rommel-Str. 1, D 91058 Erlangen, Germany \and
University of Namibia, Department of Physics, Private Bag 13301, Windhoek, Namibia \and
University of Durham, Department of Physics, South Road, Durham DH1 3LE, U.K. \and
DESY, D-15738 Zeuthen, Germany \and
Institut f\"ur Physik und Astronomie, Universit\"at Potsdam,  Karl-Liebknecht-Strasse 24/25, D 14476 Potsdam, Germany
% ...
}

\newpage

\offprints{\\J\'er\'emie M\'ehault: mehault@cenbg.in2p3.fr\\Marie-H\'el\`ene Grondin: grondin@cenbg.in2p3.fr}


\abstract
{}
{Previous observations with the High Energy Stereoscopic System (H.E.S.S.) have revealed the existence of an extended very-high-energy (VHE; E$>$100 GeV) \g-ray source, \object{HESS J1834$-$087}, coincident with the supernova remnant (SNR) W41.
The origin of the \g-ray emission has been further investigated with the H.E.S.S. array and the Large Area Telescope (LAT)  on board the \emph{Fermi Gamma-ray Space Telescope (Fermi)}.}
{The \g-ray data provided by 61 hours of observations with H.E.S.S. and 4 years with the \emph{Fermi} LAT have been analyzed, covering over 5 decades in energy from 1.8 GeV up to 30 TeV.
 The morphology and spectrum of the TeV and GeV sources have been studied and multi-wavelength data have been used to investigate the origin of the \g-ray emission towards W41.}
{The TeV source can be modeled with a sum of two components: one point-like and one significantly extended ($\sigma_{\textrm{TeV}}=0.17^\circ\pm0.01^\circ$), both centered on SNR W41 and exhibiting spectra described by a power law with index $\Gamma_{\textrm{TeV}}\simeq2.6$.
The GeV source detected with \emph{Fermi} LAT is extended ($\sigma_{\textrm{GeV}}=0.15^\circ\pm0.03^\circ$) and morphologically matches the VHE emission.
Its spectrum can be described by a power-law model with an index $\Gamma_{\textrm{GeV}}=2.15\pm0.12$ and joins smoothly the one of the whole TeV source. A break appears in the \g-ray spectra around 100 GeV.
No pulsations have been found in the GeV range.
}
{Two main scenarios are proposed to explain the observed emission: a pulsar wind nebula (PWN) or the interaction of SNR W41 with an associated molecular cloud.
X-ray observations suggest the presence of a point-like source (a pulsar candidate) near the center of the remnant and non-thermal X-ray diffuse emission which could arise from the possibly associated PWN.
The PWN scenario is supported by the compatible positions of the TeV and GeV sources with the putative pulsar.
However, the spectral energy distribution from radio to \g-rays is reproduced by a one-zone leptonic model only if an excess of low-energy electrons is injected following a Maxwellian distribution by a pulsar with a high spin-down power ($> 10^{37}$~erg~s$^{-1}$). This additional low-energy component is not needed if we consider that the point-like TeV source is unrelated to the extended GeV and TeV sources.
The interacting SNR scenario is supported by the spatial coincidence between the \g-ray sources, the detection of OH (1720 MHz) maser lines and the hadronic modeling.
}

\authorrunning{H.E.S.S. Collaboration}
\titlerunning{HESS J1834--087 studied with H.E.S.S. and {\it Fermi} LAT}
\maketitle

\section{Introduction}\label{Sec:Intro}
	During the 2005-2006 Galactic Plane Survey in the very-high-energy (VHE; E$>100$ GeV) range, the High Energy Stereoscopic System (H.E.S.S.), an array of imaging atmospheric Cherenkov telescopes, revealed more than a dozen new sources \cite[]{galsurvey1, HESSJ1834_HESS}.
	Extensions of the survey and deeper observations of the Galactic plane led to the detection of more than 80 sources \citep{HessPlaneSurvey}.
	While many Galactic TeV \g-ray sources can be identified with counterparts at other wavelengths, such as the binary LS 5039 \citep{ls1, ls2}, the supernova remnant (SNR) RX J1713$-$3946 \citep{RXJ1713_HESS} or HESS J1356$-$645 \citep{J1356_HESS} associated with a pulsar wind nebula (PWN), more than 20 \g-ray sources remain unidentified to date.
	Pulsar wind nebulae are the dominant class of Galactic TeV sources with at least 27 identifications up to now out of a total of more than 35 sources considered to be potentially due to PWN emission.
	The presence of an energetic pulsar close to the position of the source is an important clue in the identification process. 

	\object{HESS J1834$-$087} is one such unidentified source \citep{HESSJ1834_HESS}, later detected by the MAGIC telescope \citep{HESSJ1834_MAGIC}.
	It features bright and extended (intrinsic Gaussian width $\sigma=0.2^\circ$) VHE emission spatially coincident with the SNR G23.3$-$0.3 (W41).
	Additionally, high-energy (HE; E $>$ 100 MeV) emission spatially coincident with W41 was detected with the \emph{Fermi} Large Area Telescope (LAT) and listed in the first- and second-year catalogs as 1FGL J1834.3$-$0842c \citep{FstLatSrcList} and 2FGL J1834.3$-$0848 \citep{ScdLatSrclist}, respectively.

	SNR W41 shows an incomplete shell of $\sim33^\prime$ diameter in the radio domain \citep{AgeW41}.
	Several H{\scriptsize II} regions are spatially coincident with the shell but unrelated to the remnant as explained by \cite{LastDistW41}, who also estimated the kinematic distance of W41 to be between 3.9 and 4.5 kpc based on H\textrm{I} and CO observations.
	The associated cloud has a radial velocity in the local standard of rest (\emph{lsr}) of V$_{Cl_{lsr}} = 77$ km s$^{-1}$.
	\cite{OHmaserW41} detected OH (1720 MHz) maser line emission, which demonstrates the physical association between the SNR and the molecular cloud (MC).
	The maser is located near the center of the remnant and its radial velocity (V$_{\textrm{OH}_{lsr}} \simeq 74$ km s$^{-1}$) coincides with the MC velocity.
	\cite{AgeW41} estimated the ambient density $n\sim6$ cm$^{-3}$, and, by applying a Sedov model \citep{AgeSNR}, derived an age of $6\times10^4$ yr. However, the observed radius of the shock front is larger than the radius of the shell merged with the surrounding medium (called complete cooling in \cite{AgeSNR}). The age determined by \cite{AgeW41} assuming a rapid cooling of the blast wave and a completed shell formation is $\sim2\times10^5$ yr.
	
	In X-rays, \cite{XMMObs}, with the \emph{XMM-Newton} observatory, detected a central compact object (CCO; XMMU J183435.3$-$084443) at the center of the remnant  ($l=23.236^\circ, b=-0.270^\circ$) and a faint tail-like emission.
	The spectral parameters of this diffuse emission, which could be a PWN, are extracted in a region of 0.1\d~radius.
	The integrated energy flux and the spectral index are $F_{2-10~\textrm{keV}}=(4.2\pm2.2)\times10^{-13}$ erg cm$^{-2}$ s$^{-1}$ and $\Gamma=1.9\pm1.0$, respectively.
	Perhaps due to a better resolution with respect to the European Photon Imaging Camera (EPIC) onboard \emph{XMM-Newton} and a short observation time ($\sim40$ ks), \cite{W41Chandra} did not report the offset emission in \emph{Chandra} data but a compact non-thermal X-ray nebula surrounding the CCO (CXOU J183434.9$-$084443), which could be a dust halo or a PWN.
	Despite the non-detection of X-ray pulsations from the CCO, the existence of the non-thermal X-ray nebula suggests that it could be a pulsar.
	
	No radio emission from the putative pulsar or its nebula has been detected so far. Radio upper limits were derived from the Very Large Array (VLA) observations at 20 cm \citep{VLASurvey} and the 1.1 mm Bolocam \citep{BoloCam1mm} Galactic Plane Survey images within a 0.1\d~radius around the position of the point-like X-ray source.

	Two scenarios have been proposed to explain the TeV emission: the interaction between the SNR W41 and a nearby MC \citep{LiChenW41Had,CastroW41} and a PWN powered by the compact object CXOU J183434.9$-$084443 \citep{W41Chandra}.

	The analyses presented in this work have been performed using data taken with the H.E.S.S. telescopes and \emph{Fermi} LAT.
	H.E.S.S. is an array of four identical 13-m diameter imaging telescopes located in the Khomas Highland of Namibia at an altitude of 1800 m. A larger telescope with 28-m diameter (H.E.S.S. Phase II) has been built in the center of the array and saw its first light in July 2012.
	No data from H.E.S.S. Phase II have been used in this work.
	H.E.S.S. is sensitive to \g-rays at energies above 100 GeV up to several tens of TeV and has a large field-of-view (FoV) of 5$^{\rm o}$ diameter.
	Requiring at least two telescopes triggered by an air shower, the system yields a very good angular resolution ($r_{68\%}\sim0.1$\d), energy resolution ($\Delta E/E\sim 15 \%$), and an efficient background rejection \citep{PresentHess}. These characteristics, together with its geographic location, make H.E.S.S. ideally suited for discoveries and studies of extended sources in the Galactic plane.\\
	The LAT on board the \emph{Fermi Gamma-ray Space Telescope (Fermi)}, launched in June 2008, is a pair conversion telescope sensitive to photons in a broad energy band from $\sim$ 20 MeV to more than 300 GeV.
	It has a wide FoV of 2.4 sr and a good angular resolution ($r_{68\%}\lesssim1^\circ$ above 1 GeV).
	For a detailed description of the instrument, see \cite{DescLat}.
	
	Deeper observations of the region of SNR W41 with H.E.S.S. and \emph{Fermi} LAT since the discovery of the \g-ray source allow morphological and spectral analyses over a very wide energy range of a few GeV to several tens of TeV, providing new insights on the origin of the \g-ray emission coincident with SNR W41.\\
	The H.E.S.S. observations, data analysis and the characteristics of HESS J1834$-$087 are provided in Section \ref{Sec:HessAnal}, while the results from the analysis of the \emph{Fermi} LAT data are presented in Section \ref{Sec:LatAnal}. A discussion on possible \g-ray emission mechanisms is given in Section \ref{Sec:Discuss}.

\section{H.E.S.S. observations and data analysis}\label{Sec:HessAnal}
	H.E.S.S. observations of 28 min duration were taken in wobble mode centered on the position of the SNR W41. The pointing alternates between offsets of $\pm$0.5\d~in declination and right ascension with respect to the nominal target position.
	After standard quality selection \citep{PresentHess} to remove data affected by unstable weather conditions or hardware-related problems, the total live-time of the data set, including data presented in the discovery paper \citep{HESSJ1834_HESS}, is 61 hours.
	The zenith angle of the observations ranges from 10\d~to 45\d, with a median value of 20\d.

	The data were analyzed with the Model Analysis method, implemented in the ParisAnalysis software (version 0-8-22), which compares the raw atmospheric shower images in the Cherenkov camera with the prediction of a semi-analytical model \citep{Modpp}. Two different event selection cuts have been applied.	
	Standard cuts, which require a minimum shower image intensity of 60 photo-electrons (p.e.) in each camera, were used to compute energy spectra and sky maps.
	To improve the angular resolution and obtain better background rejection, hard cuts (higher minimum intensity of 120 p.e.) were used for morphological studies.
	The energy thresholds for standard and hard cuts are 177 GeV and 217 GeV, respectively.
	A second analysis \citep[HAP, ][]{PresentHess}, which includes independent calibration of pixel amplitude and event reconstructions and discrimination based on Hillas moments of the cleaned shower image, has been used for cross-checks and provides compatible results. 

\begin{figure}[t!]
        \centering
        \includegraphics[width=0.5\textwidth]{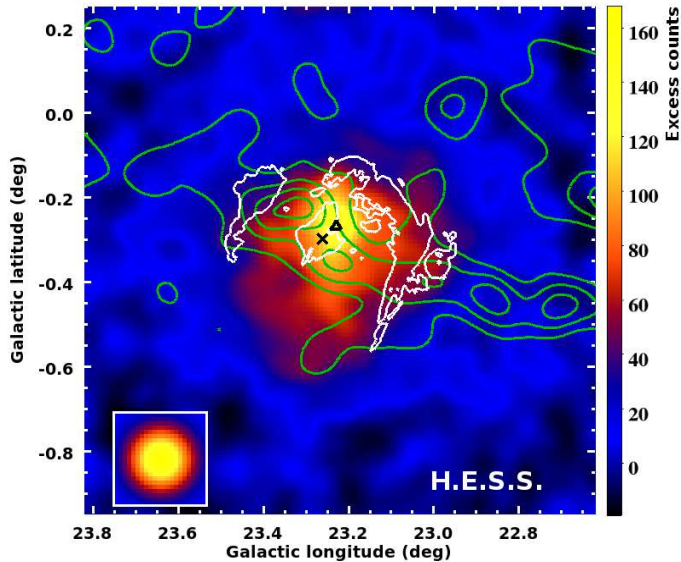}
        \caption{H.E.S.S. image of the VHE \g-ray excess in the direction of SNR W41 after standard cuts analysis. The image was smoothed with a Gaussian kernel of $0.068$\d~corresponding to the instrument's point-spread function for this data set, which is shown in the inset. Green contours represent the GRS $^{13}$CO data around W41 integrated in the velocity range 74 km s$^{-1}$ to 82 km s$^{-1}$ and smoothed with the same kernel. White contours of the VLA radio data at 20 cm show the shell-type SNR. The position of CXOU J183434.9$-$084443 and the brightest OH (1720 MHz) maser are indicated with a triangle and a cross, respectively. The linear color scale is in units of counts per smoothing Gaussian width.}
        \label{Fig:W41AllE}
\end{figure}

	\subsection{Morphological analysis}
		\label{Sec:HessMorpho}
\begin{figure}[t!]
			\centering
 			\includegraphics[width=0.5\textwidth]{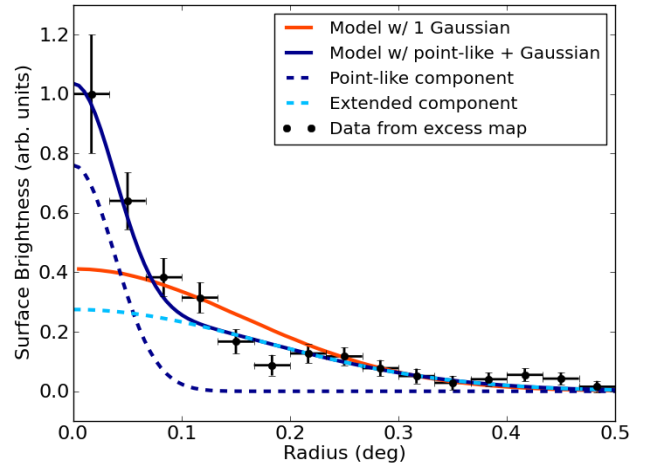}
			\caption{H.E.S.S. radial profile of the uncorrelated excess map centered at the position of the putative pulsar. The solid dark blue line shows the result of the best-fit model (labelled (C) in Table \ref{Tab:TeVMorpho}) estimated in Section \ref{Sec:HessMorpho}. The dark blue  and light blue dashed lines represent the point-like and Gaussian components of this model. The profile from the single Gaussian model (labelled (B) in Table \ref{Tab:TeVMorpho}) is plotted in orange.}
			\label{Fig:RadProf}
		\end{figure}

\begin{figure}[t!]
			\centering
			\includegraphics[width=0.5\textwidth]{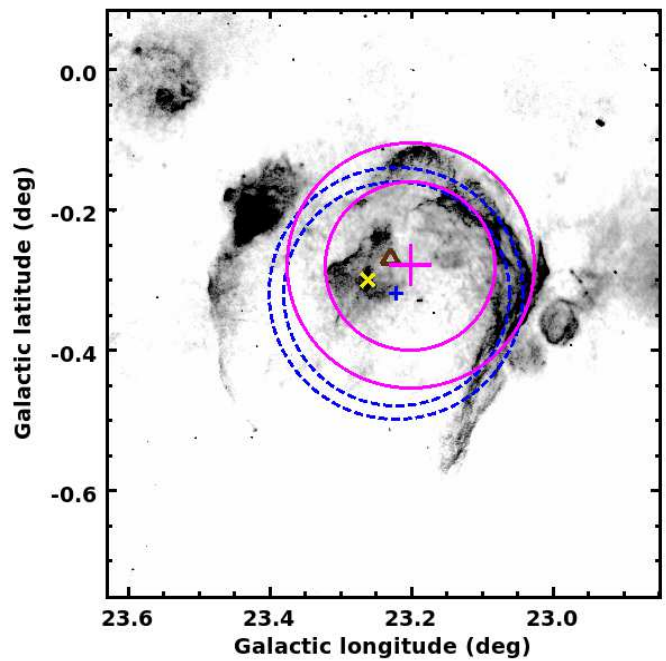}
			\caption{20 cm radio continuum map \citep{VLASurvey} with superimposed best-fit positions as vertical crosses (their sizes indicate the statistical errors) and intrinsic extensions as rings. Magenta and dashed blue represent respectively the Gaussian model of the \emph{Fermi} LAT source and the extended component of the 2-component model of the H.E.S.S. source. The inner and outer radius represents $\pm1~\sigma$ errors on the fitted extension. The candidate pulsar (triangle) and OH maser (cross) positions are reported in brown and yellow, respectively.}
			\label{Fig:GeVTeVMorpho}
		\end{figure}

		Figure \ref{Fig:W41AllE} presents the H.E.S.S. excess map obtained after the analysis with standard cuts and smoothed with a Gaussian kernel of $0.068$\d~which corresponds to the mean point-spread function (PSF) for this data set.
		Cloud density traced by $^{13}$CO from the Galactic Ring Survey \cite[GRS;] []{GRSData} smoothed with the same Gaussian kernel, radio 20 cm contours \citep{Vla20cm} from W41, and the position of the CCO CXOU J183434.9$-$084443 are shown as well.\\
		The detection of the TeV source is clear, with a significance of $\simeq27~\sigma$ within an integration radius of $0.3^\circ$ centered on the candidate pulsar position.
		The following morphological results have been obtained with the tools included in ParisAnalysis software.
		The best-fit position assuming a point-like model (model (A) in Table \ref{Tab:TeVMorpho}) can be considered compatible with the pulsar position (0.01\d~away from it).
		Using a symmetrical Gaussian model (model (B) in Table \ref{Tab:TeVMorpho}) significantly improves the likelihood of the fit with respect to the previous model. The best-fit position moves 0.04\d~away from the putative pulsar but is still compatible with it.
		The sum of a point-like and a symmetrical Gaussian convolved with the PSF (model (C) in Table \ref{Tab:TeVMorpho}) again improves ($5~\sigma$) the fit with respect to the single Gaussian model. No significant extension of the central component has been found.
		As can be seen in Figure \ref{Fig:RadProf}, the radial profiles of the unsmoothed best-fit model (model (C)) and the excess map centered at the position of the putative pulsar are in good agreement. For comparison, the radial profile of the single Gaussian model (model (B)) has been displayed.
		Table \ref{Tab:TeVMorpho} summarizes the tested models and Figure \ref{Fig:GeVTeVMorpho} represents the best-fit morphology.
		
		\begin{table*}[htbp]
		\begin{center}
            \begin{minipage}[t]{0.9\textwidth}
			\begin{center}
                        \caption{Centroid (Galactic coordinates) and extension fits to the H.E.S.S. events spatial distribution in the W41 region. The errors given are statistical only. The test statistic (TS) (defined in Section \ref{Sec:FermiMorpho}) quantifies the improvement of a model to a more complex one.}
                        \label{Tab:TeVMorpho}
				\begin{tabular}{r l c c c c c c c }
			\hline
			\hline
		~ &Model tested & TS & $S_{\textrm{improv}}$\footnote{Significance of the improvement for each of the alternative models with respect to the previous less complex model.} 								& $N_\textrm{par}$\footnote{Number of free parameters.} & $l$ [	$^\circ$]  &       $b$ [$^\circ$]   &  $\sigma$ [$^\circ$]\footnote{$\sigma$ is the intrinsic size of the source, i.e. PSF substracted.}\\
			\hline
		$(A)$ & Point-like & - &    -    & 3 &  $23.222\pm0.004$  &  $-0.268\pm0.004$ & --  \\
			\hline
		$(B)$ & Gaussian &  566.4 & 23.8 & 4 &  $23.22\pm0.01$  & $-0.31\pm0.01$  & $0.15\pm0.01$\\
			\hline
		~ & Point-like & \multirow{3}{*}{31.6}&\multirow{3}{*}{5.0}&\multirow{3}{*}{7}& $23.23\pm0.01$  & $-0.26\pm0.01$  & -- \\ 
		$(C)$ &~~~~~~+& & & & & & \\
		~ &Gaussian			&     &  &  & $23.22\pm0.01$&  $-0.32\pm0.01$& $0.17\pm0.01$\\
		\hline
			\end{tabular}
			\end{center}
		\renewcommand{\footnoterule}{}
		\end{minipage}
		\end{center}
		\end{table*}
		
\begin{figure}[t!]
			\centering
			\includegraphics[width=0.5\textwidth]{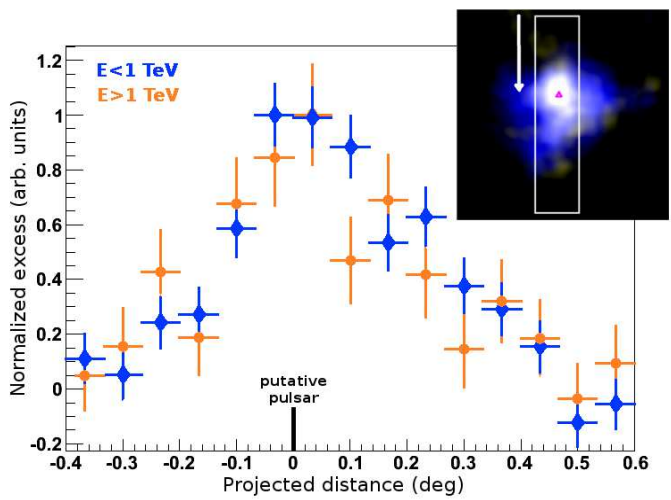}
			\caption{H.E.S.S. slices in the uncorrelated excess map along the direction perpendicular to the Galactic plane (white arrow in the inset) in two independent energy bands ($E<1$ TeV: blue and $E>1$ TeV: orange). The slices are centered at the candidate pulsar position (marked by a triangle in the inset). The composite color image (inset) shows the box ($height = 1.0^\circ, width = 0.22^\circ$) used for the profile. }
			\label{Fig:TeVBands}
		\end{figure}

		The large energy range covered by H.E.S.S. allows a morphological study in two energy bands: we have used data below and above 1 TeV to get equivalent statistics in each band.
		Slices perpendicular to the Galactic plane do not show any significant variation of the morphology as seen in Figure \ref{Fig:TeVBands}. The same conclusion has been obtained along other angles.
		The corresponding composite image (Figure \ref{Fig:TeVBands}, inset) highlights that the maximum emission remains centered at the putative pulsar position.
		
		\subsection{Spectral analysis}
			\label{Sec:HessSpec}
			Spectral information has been obtained  between 0.2 and 30 TeV in two distinct regions corresponding to the best morphological model.
			The central and annular emissions were analyzed within a circular region of $0.1^\circ$ radius and between $r_{\textrm{min}}=0.1^\circ$ and $r_{\textrm{max}}=0.3^\circ$ respectively. 
			The \emph{reflected region background} technique \citep{TeVBkgTechn} has been used to evaluate the background in the FoV.
			The spectrum of the central region is well fitted by a pure power law (equivalent $\chi^2/ndf = 22/18$) defined as:
				\begin{eqnarray}
						\frac{\textrm{d}N}{\textrm{d}E} = \Phi(E_\textrm{dec})\times \left(\frac{E}{E_\textrm{dec}}\right)^{-\Gamma}
						\label{Eq:PowLaw}
				\end{eqnarray}
			where the decorrelation energy $E_\textrm{dec}$ is the energy at which the correlation between the flux normalization ($\Phi$) and spectral index $\Gamma$ is zero and where the statistical errors are the smallest.
			The best-fit parameters, listed in Table \ref{Tab:TeVSpect}, yield an integrated flux $I_{1-30~\textrm{TeV}}=(0.40\pm0.05_{\textrm{stat}}\pm0.08_{\textrm{syst}})\times10^{-12}$ cm$^{-2}$ s$^{-1}$, which corresponds to $\sim2\%$ of the Crab Nebula flux in the same energy range.
			The use of an exponential cutoff power-law model improves the fit at a $2.4~\sigma$ level, slightly less than the improvement level reported in \cite{Icrc11W41}, but still in agreement with it.
			In the following, the pure power-law shape is considered.

\begin{figure}[t!]
				\centering
				\includegraphics[width=0.5\textwidth]{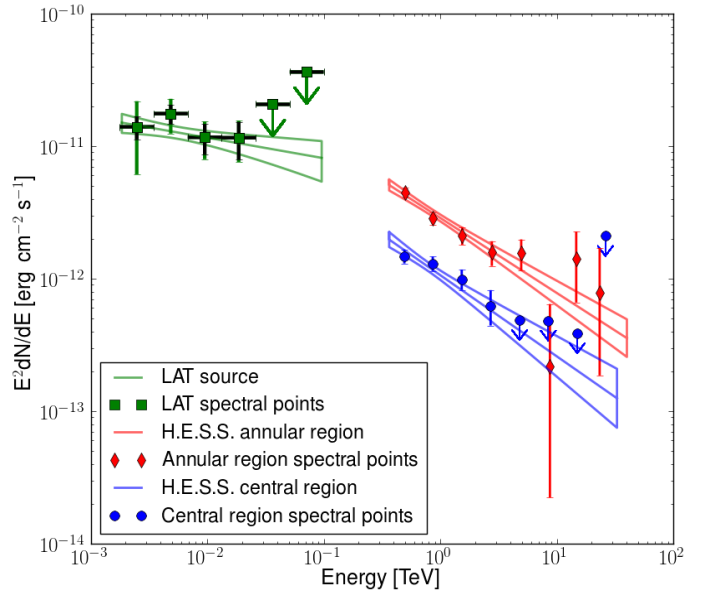}
				\caption{Spectral points and bowties of the LAT source (squares - statistical and systematic uncertainties added in quadrature in green and only statistical errors in black) and the H.E.S.S. annular and central regions (respectively in red diamonds and blue points). Bowties represent 1-$\sigma$ confidence level error on spectra. The upper limits are calculated for a $3~\sigma$ confidence level.}
				\label{Fig:GeVTeVSpectra}
			\end{figure}

			The spectrum extracted from the annular region follows a power-law shape ($\chi^2/ndf = 37/32$), with best-fit parameters reported in Table \ref{Tab:TeVSpect}.
			The integrated flux $I_{1-30~\textrm{TeV}}=(1.13\pm0.11_{\textrm{stat}}\pm0.23_{\textrm{syst}})\times10^{-12}$ cm$^{-2}$ s$^{-1}$ corresponds to $\sim5\%$ of the Crab Nebula.
			No significant difference between the spectral indices in the central and the annular region has been detected.
			The sum of the integral fluxes calculated in each region is in agreement with the one found in the total emission, the latter being well fitted by a power-law model (parameters listed in Table \ref{Tab:TeVSpect}).
			Figure \ref{Fig:GeVTeVSpectra} shows the H.E.S.S. spectral points for the annular and central regions and $1~\sigma$ statistical errors. 
			The bowties represent the $1~\sigma$ error contours of the spectra.
			The measured fluxes and spectral indices for these regions given in Table \ref{Tab:TeVSpect} and Figure \ref{Fig:GeVTeVSpectra} are reliable, because these regions are large compared to the H.E.S.S. PSF and the H.E.S.S. PSF shows little variation with energy.
			\begin{table*}[htbp]
			\begin{center}
				\begin{minipage}[c]{0.9\linewidth}
				\begin{center}
					\caption{H.E.S.S. best-fit spectral parameters assuming a power-law model (Equation \ref{Eq:PowLaw}) between 0.2 and 30 TeV. The systematic errors are conservatively estimated to be $\pm0.2$ on the photon index and $20\%$ on the flux \citep{PresentHess}.}
					\label{Tab:TeVSpect}
					\begin{tabular}{lccccc}
			\hline
			\hline
		Region & $\Phi(E_\textrm{dec})$ & $E_\textrm{dec}$ &  $\Gamma$ & equivalent $\chi^2/ndf$ & $I_{1-30~\textrm{TeV}}$ \\
			& $\times10^{-12}$ [cm$^{-2}$ s$^{-1}$ TeV$^{-1}$] & [TeV]& &    &$\times10^{-12}$ [ph cm$^{-2}$ s$^{-1}$] \\
			\hline
		Central & 2.28$\pm$0.15$_{\textrm{stat}}$ & 0.633 & 2.67$\pm$0.11$_{\textrm{stat}}$ & 22/18 & 0.40$\pm$0.05$_{\textrm{stat}}$ \\
			\hline
		Annular & 4.33$\pm$0.22$_{\textrm{stat}}$ & 0.716 & 2.60$\pm$0.07$_{\textrm{stat}}$ & 37/32 & 1.13$\pm$0.11$_{\textrm{stat}}$ \\
			\hline
		Total   & 6.50$\pm$0.27$_{\textrm{stat}}$ & 0.687 & 2.64$\pm$0.06$_{\textrm{stat}}$         & 29/32 & 1.46$\pm$0.11$_{\textrm{stat}}$ \\
					\hline
					\end{tabular}
				\end{center}
				\end{minipage}
                \end{center}
                 \end{table*}
			
			As explained in Section \ref{Sec:HessMorpho}, the VHE \g-ray emission is modeled by the sum of two components: one point-like and one extended.
			In order to estimate the intrinsic flux of each component and to model the spectral energy distribution (SED), the spectral points extracted in the central and annular regions (see Figure \ref{Fig:GeVTeVSpectra}) have to be corrected for the contributions of each component in each region.
			To normalize only their flux, a constant spectral index at any distance from the putative pulsar has been assumed in agreement with the spectral fits (see Table \ref{Tab:TeVSpect}).
			From the best morphological model, the contributions of the point-like and extended components to the total flux have been respectively calculated to be $64\%\pm4\%$ and $36\%\pm4\%$ of the central region flux ($r<0.1^\circ$). The point-like component contribution to the total flux emission in the annular region is $17\%\pm2\%$.

\section{\emph{Fermi} LAT observations and analysis}\label{Sec:LatAnal}

\begin{figure}[t!]
        \centering
        \includegraphics[width=0.5\textwidth]{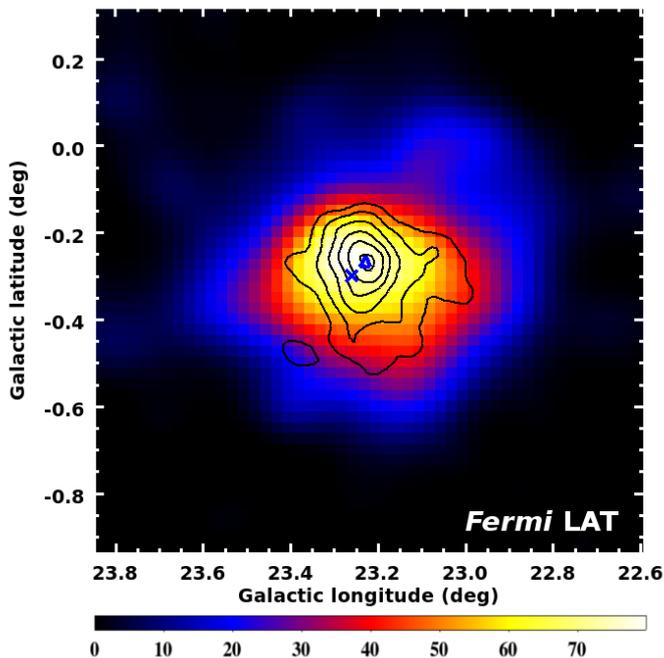}
        \caption{\emph{Fermi} LAT TS map of the W41 region, above 1.8 GeV. The TS was evaluated by placing a point-source at the center of each pixel, Galactic diffuse emission and nearby 2FGL sources are included in the background model. H.E.S.S. significance contours (3, 4, 5, 6, 7 and 8$~\sigma$) are overlaid as black solid lines. The triangle and the cross indicate the position of the putative pulsar and the OH maser respectively.}
        \label{Fig:Fermi}
\end{figure}

	The following analysis was performed using 47 months of data acquired from 2008 August 4 to 2012 June 30 within a 15\d~$\times$ 15\d~square around the position of the SNR W41.
	Only \g-ray events with reconstructed zenith angles smaller than 100$^{\circ}$ have been selected in order to reduce contamination from \g~rays from cosmic-ray interactions in the upper atmosphere.
	In addition, time intervals have been excluded when the rocking angle is more than 52\d~and when the \emph{Fermi} satellite is in the South Atlantic Anomaly.
	Events have been selected from 1.8 GeV to 100 GeV in the Pass 7 Source class \citep{P7Paper}. The \emph{P7\_SOURCE\_V6} instrument response functions (IRFs) have been used.
	This selection is motivated by both the morphological stability of the source, the improved angular resolution at high energy, and the difficulties in modeling the Galactic diffuse emission implying large systematic errors at low energy. These effects can be seen in the low-energy part ($E<1$ GeV) of the spectrum in \cite{CastroW41}.
	Two different tools using the maximum likelihood technique \citep{mattox96} were applied to perform the spatial and spectral analysis: $\mathtt{pointlike}$ \citep{PointLikePap} and \emph{Fermi} LAT Science Tools \footnote{\texttt{\tiny http://fermi.gsfc.nasa.gov/ssc/}}.
	These tools fit a source model to the data along with models for the residual charged particles, diffuse $\gamma$-ray emission and the LAT sources in the region of interest listed in the second LAT catalog \citep{ScdLatSrclist}.
	To describe the Galactic diffuse emission, the ring-hybrid model {\it gal\_2yearp7v6\_v0.fits} has been used.
	The instrumental background and the extragalactic radiation are described by a single isotropic component with the spectral shape in the tabulated model {\it iso\_p7v6source.txt}.
	These models are available from the \emph{Fermi} Science Support Center\footnote{\texttt{\tiny http://fermi.gsfc.nasa.gov/ssc/data/access/lat/\\BackgroundModels.html}}.

	All 2FGL sources within 20\d~radius around W41 were added to the spectral-spatial model of the region.
	\cite{PointLikePap} performed a detailed study of the region surrounding HESS J1837$-$069 ($l\sim25.3^\circ;b\sim-0.3^\circ$) based on the LAT catalog.
	The best model (including spectrum) describing this region has been used. The source 2FGL J1837.3$-$0700c is modeled with a symmetrical Gaussian ($\sigma\simeq0.33^\circ$). The two closest sources, 2FGL J1834.7$-$0705c and 2FGL J1836.8$-$0623c, are relocalized even if their positions did not significantly change.

	\subsection{Search for pulsed emission}
		The \emph{Fermi} LAT is the first instrument sensitive enough to discover new pulsars based on their \g-ray pulsations alone \citep{LatBlindSrch}: 41 new \g-ray pulsars have been detected to date in this way.
		To check for the presence of a \g-ray pulsar in W41, a search for pulsations in \emph{Fermi} LAT data using the hierarchical blind search technique described in \cite{BlindSrch} has been performed.
		To improve the signal-to-noise ($S/B$) ratio of the putative pulsar in W41, each photon was assigned a probability that it originates from the target source, using the spectral model obtained from the analysis described in Section \ref{Sec:LatSpec}. For W41 $S/B\leq1$ while the faintest pulsars detected in blind searches have $S/B\sim15$ : no significant pulsations were found.

	\subsection{Morphological analysis}\label{Sec:FermiMorpho}
		Figure \ref{Fig:Fermi} shows the test statistic (TS) map between 1.8 and 100 GeV, with H.E.S.S. significance contours (from 3 to 8$~\sigma$) overlaid.
		The TS is defined as twice the difference between the log-likelihood $L_1$ obtained by fitting a source model plus the background model to the data, and the log-likelihood $L_0$ obtained by fitting the background model only, \emph{i.e} TS$=2\ln(L_1-L_0)$. 
		This skymap contains the TS value for a point source at each map location, thus giving a measure of the statistical significance for the detection of a \g-ray source in excess of the background.
		The source is detected with TS$_{\textrm{max}}=111$  above 1.8 GeV.
		During the fit, the spectral parameters of sources closer than 5$^{\circ}$ to W41 were allowed to vary in the likelihood fit as well as the normalization of the diffuse models, while the parameters of all other sources were fixed at the values of the 2FGL catalog.
		The source extension has been determined using $\mathtt{pointlike}$ with a Gaussian hypothesis (compared to a point-source hypothesis).
		The difference in TS between the Gaussian and the point-source hypothesis is 20, which converts to a significance of $\sim 4.5~\sigma$.
		This suggests that the source is extended with respect to the LAT PSF.
		The intrinsic size is $\sigma_{\textrm{GeV}}=(0.15^\circ\pm0.03^{\circ})$ and the best-fit position $(l=23.20^\circ\pm0.03^\circ$, $b=-0.28^\circ\pm0.03^\circ)$ is compatible with the candidate pulsar position.
		This morphology is shown in Figure \ref{Fig:GeVTeVMorpho}.
		An asymmetric Gaussian distribution was also considered but did not improve the fit with respect to the Gaussian hypothesis ($\Delta$TS=1.5).

	\subsection{Spectral analysis}
	\label{Sec:LatSpec}
		The spectrum of the \emph{Fermi} LAT source has been estimated with $\mathtt{gtlike}$ taking into account the sky model and the shape obtained with $\mathtt{pointlike}$. 
		The spectrum is well described by a power law (see Equation \ref{Eq:PowLaw}).
		The best-fit parameters are: $\Phi(E_\textrm{dec})=(4.98\pm0.20_{\textrm{stat}}\pm0.22_{\textrm{syst}})\times10^{-13}$ cm$^{-2}$ s$^{-1}$ MeV$^{-1}$, $E_\textrm{dec} = 4088$ MeV and $\Gamma=2.15\pm0.12_{\textrm{stat}}\pm0.16_{\textrm{syst}}$.
		These results are compatible with the parameters found by \cite{CastroW41}.

		Two main sources of systematic errors have been considered: imperfect modeling of the Galactic diffuse emission and uncertainties in the effective area.
		The first one was estimated by changing artificially the normalization of the Galactic diffuse model by $\pm6\%$ as done in \cite{W49B_Fermi}.
		The second one is estimated by using modified IRFs whose effective areas bracket the nominal ones.
		These bracketing IRFs are defined by envelopes above and below the nominal energy dependence of the effective area by linearly connecting differences of ($10\%$, $5\%$, $20\%$) at log(E/MeV) of (2, 2.75, 4), respectively \citep{P7Paper}.

		The \emph{Fermi} LAT spectral points were obtained by dividing the energy range (1.8 $-$ 100 GeV) in 6 logarithmically-spaced energy bins.
		A Bayesian upper limit \citep{BayesUl} at $3~\sigma$ confidence level is calculated if the TS value in the energy bin is lower than 9.
		The LAT spectral points and their uncertainties are reported in Figure \ref{Fig:GeVTeVSpectra}.

		The scenario where the emission seen with \emph{Fermi} LAT is produced by two components has been investigated by fixing the best-fit extension and position found in GeV and adding a point-like source at the putative pulsar position.
		In this case, the point-like source is not significant ($\textrm{TS}\simeq7$) and a Bayesian upper limit on its integrated flux has been calculated: $S(1.8-100~\textrm{GeV})<1.0\times10^{-12}$ erg cm$^{-2}$ s$^{-1}$  at $3~\sigma$ confidence level.


\section{Discussion}\label{Sec:Discuss}
	Follow-up observations of the previously discovered source HESS J1834$-$087 as described here have led to the discovery of a two-component morphology: a central point-like and an extended component.
	\emph{Fermi} LAT observations were also analyzed in the GeV energy band leading to the detection of a \g-ray source (2FGL J1834.3$-$0848) coincident with the TeV source and with a similar extension.
	Figure \ref{Fig:GeVTeVMorpho} highlights that the positions of both the TeV and GeV sources are compatible to each other and also with SNR W41.
	This new information is very useful to constrain the nature of the \g-ray source.

	In addition to the extended emission, no significant point-like component was detected using \emph{Fermi} LAT data, in contrast with H.E.S.S. (see Section \ref{Sec:HessMorpho}).
	With \emph{Fermi} LAT observations, \cite{2PC_Fermi} have found that pulsar spectra usually follow a power law with an exponential cutoff at energies between 1 and 10 GeV.
	However, such a cutoff is not apparent in the spectrum derived in Section \ref{Sec:LatSpec}.
	Moreover, no \g-ray pulsations from the putative pulsar at the center of the remnant have been detected.
	These arguments and the fact that the \g-ray emission observed with \emph{Fermi} LAT is significantly extended, strongly suggest that the signal does not arise from an unseen pulsar.

	It is clear from Figure \ref{Fig:W41AllE} that the VHE \g-ray emission does not correlate with the SNR radio shell, but comes from the center of the remnant as shown by the VHE radial profile in Figure \ref{Fig:RadProf}.
	This disfavors a scenario in which the signal originates in the SNR shell in contrast with the case of, $e.g.$, RX J1713.7$-$3946 \citep{RXJ1713_HESS}.

	Therefore, only two plausible scenarios remain: either the \g-ray emission is produced by a PWN powered by the pulsar candidate CXOU J183434.9$-$084443 detected in X-rays or it is due to the interaction of the SNR with a nearby MC.
	
	Unless otherwise specified, an assumed distance of $d=4.2$ kpc will be considered in the following sections.

		\subsection{Pulsar wind nebula candidate scenario}
			The increasing number of detected PWNe at VHE motivates the investigation of a scenario where the GeV and TeV \g-ray emission is produced by the wind of the putative pulsar CXOU J183434.9$-$084443.
			\subsubsection{TeV source HESS J1834$-$087 as a single source powered by the putative pulsar}
			The spatial coincidence between the centroid of the GeV and the TeV emissions and the pulsar candidate strengthens this hypothesis, as also stated by \cite{W41Chandra}.
			Interestingly, assuming that the extended component detected in X-rays is a PWN, \cite{W41Chandra} found that the ratio of the extended source and pulsar candidate X-ray luminosities, $L_{\textrm{PWN}_X}$/$L_{\textrm{PSR}_X} \sim 1.8$, is typical of a PWN/pulsar system.
			Furthermore, the PWN scenario is reasonable from an energetic point of view.
			Indeed, assuming a distance of $d=4.2$ kpc, \cite{W41Chandra} derived an X-ray luminosity of $L_X(2 - 10~\textrm{keV})\simeq1\times10^{33}$ erg s$^{-1}$. Then, assuming a termination shock radius $r_{s, 17} = r_s / (10^{17}$ cm) scaled to a plausible value (corresponding to an angular size of 2.7'' at the assumed distance) 
and an ambient pressure $p_{amb, -9} = p_{amb}/ (10^{-9}$ dyne~cm$^{-2}$) inside the SNR in the Sedov expansion phase, the authors derive a spin-down luminosity of $\dot E = 4 \times 10^{36} \times r^2_{s, 17} \times p_{amb, -9}$ erg s$^{-1}$, which is typical for Vela-like pulsars and supports the assumption of a young PWN. This value is consistent with the estimates of  $[10^{36}, 10^{37}]$ erg s$^{-1}$ that can be derived from \cite{Kargaltsev2008}.
			Pulsars associated with several TeV PWNe have a similar spin-down power \citep{HessPwn}.
			
			From the data analyses described in the previous sections, the GeV and TeV luminosities are respectively estimated to be $L_{\textrm{GeV}}(1 - 100~\textrm{GeV})\simeq1.6\times10^{35}$ erg s$^{-1}$ and $L_{\textrm{TeV}}(1 - 30~\textrm{TeV})\simeq1.1\times10^{34}$ erg s$^{-1}$, the latter being calculated for the entire emitting region.
			\cite{MattanaPwn} used a large sample of PWNe observed in X-rays and \g-rays in order to demonstrate the existence of a correlation between the luminosity ratio $L_{\textrm{TeV}}/L_X$ and the characteristic age $\tau_c$ of the powering pulsar.
			Using the luminosities derived from \emph{XMM-Newton} and H.E.S.S. observations, the characteristic age is in the range $\tau_c \in [3\times10^3, 6\times10^4]$ yr, \emph{i.e} younger than the estimated age of the remnant (tentative remnant dating: $t_{\textrm{SNR}}\in[6\times10^4, 2\times10^5]$ yr ; see Section \ref{Sec:Intro}).

			At the assumed distance of 4.2~kpc, the radius of the nebula in the TeV range is $r\simeq12$ pc. 
			A corresponding age of $[6\times10^3, 2\times10^4]$ yr is obtained by modeling the PWN evolution inside the SNR \citep{ModelPwnInSnr} with a mass of the progenitor $M\sim8$ M$_\odot$, an explosion energy $E_0\sim10^{51}$ erg, a spin-down luminosity of $\dot E \in [10^{36}, 10^{37}]$ erg s$^{-1}$ and a density inside the shell $n=0.1$ cm$^{-3}$.
			By using standard values, this age is in agreement with the estimated characteristic age of the putative pulsar described above, but still lower than the interval proposed for the remnant.
			
			Therefore, if HESS J1834$-$087 is a PWN powered by the putative pulsar CXOU J183434.9$-$084443, this would imply that the SNR W41 might be younger than $6\times10^4$ yr, or that W41 is located at a distance smaller than the assumed value, or that the pulsar and the TeV source are simply not related to W41. These conclusions are strongly dependent on the input parameters and have been derived assuming an evolutionary phase before the crushing by the reverse shock supported by the X-ray observations described above. Therefore, they must be considered as indicators only.

\begin{figure}[t!]
				\centering
				\includegraphics[width=0.5\textwidth]{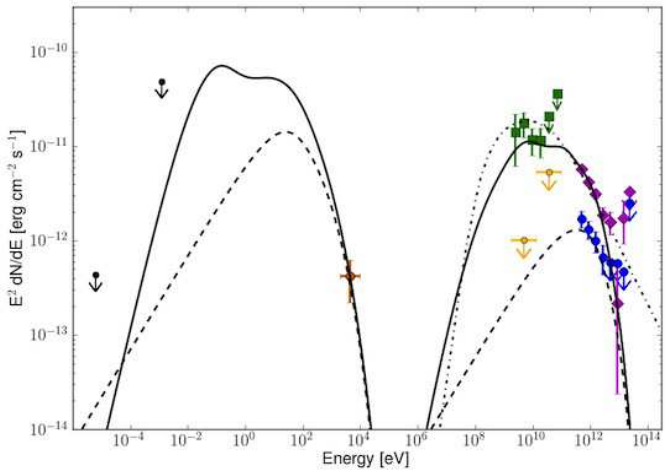}
				\caption{Spectral energy distribution (SED) for three different assumptions about the nature of the source. \emph{Solid line:} The association between a single TeV source (violet diamonds) and the GeV source (green squares) has been assumed. A leptonic scenario has been tested where a low-energy Maxwellian distribution replaces the low-energy part of the electron spectrum. The fitted parameters of this model can be found in Table \ref{Tab:ParamWholeSrc}. \emph{Dashed line:} A simple leptonic model has been tested assuming the TeV point-like source (blue points) is associated with a hypothetical GeV point-like source superimposed on the extended one (yellow upper limits). The blue TeV spectral points have been obtained after normalization of those computed in the central extracted region (see Section \ref{Sec:HessSpec} for details). \emph{Dash-dotted line:} The hadronic model studied by \cite{LiChenW41Had} has been superimposed to the new data points obtained in this work. 
				The brown X-ray spectral point and black radio upper limits have been derived respectively from the \emph{XMM-Newton} spectrum given by \cite{XMMObs}, and from VLA 20 cm \citep{VLASurvey} and Bolocam 1.1 mm \citep{BoloCam1mm} images.}
				\label{Fig:SpectGeVTeVSingle}
			\end{figure}

			An SED modeling has been performed assuming that the whole TeV source is associated with the GeV emission.
			However, unlike most PWNe detected by \emph{Fermi} LAT, which present a hard spectrum with an average spectral index $\Gamma \sim 1.8$ \citep{TeVCat}, the \emph{Fermi} LAT spectrum of HESS J1834$-$087 is rather soft.
			In this sense, this \g-ray source resembles HESS J1640$-$465 for which \cite{J1640} inferred a distinct population of low-energy electrons based on the particle-in-cell simulations performed by \cite{AccelMaxwel}.
			We have used a similar reasoning by modeling the low-energy part of the electron spectrum by an additional Maxwellian distribution. The solid black curve in Figure \ref{Fig:SpectGeVTeVSingle} represents the obtained fitted leptonic model ($\chi^2/ndf=10.5/7$). 
			The output parameters are listed in Table \ref{Tab:ParamWholeSrc} and discussed hereafter. 
			The cosmic microwave background and the Galactic interstellar radiation field at the location of the SNR W41, used to calculate the inverse Compton emission, have been derived from the estimates of \cite{porter05}.
			The energy densities for the IR and optical photon fields are respectively 1.6 eV~cm$^{-3}$ and 2.2 eV~cm$^{-3}$.
			The characteristic age of the pulsar has been assumed to be $t=10$ kyr, close to the mean of the ranges previously estimated.
			The best-fit magnetic field value ($B\simeq14~\mu$G) is similar to those found for other TeV PWNe \citep{TeVPwns}. 
			The derived total energy in electrons is $W_e\simeq2.4 \times 10^{48}$ erg.
			Assuming a constant spin-down power over the entire lifetime of the pulsar, this energy corresponds to $\sim$~700\% or $\sim$~70\% of the total energy injected by the pulsar for a spin-down power of $10^{36}$ erg~s$^{-1}$ or $10^{37}$ erg~s$^{-1}$, respectively.
			 In this context, if the whole GeV and TeV emissions correspond to the same PWN, the spin-down power of the pulsar must be close to (or even higher than) $\sim 10^{37}$ erg s$^{-1}$ currently assuming a constant energy loss rate or was much higher than $10^{36}$ erg~s$^{-1}$ in the past and decreased with time. A precise modeling of the evolution of the spin-down power and magnetic field with time is out of the scope of this paper, since most parameters for such a model are poorly constrained. 

                 \begin{table*}[htbp]\begin{center}
                 \begin{minipage}[c]{0.7\linewidth}\begin{center}
                        \caption{Fitted parameters of the SED modeling shown in Figure \ref{Fig:SpectGeVTeVSingle} assuming HESS J1834$-$087 is a single physical source as observed from the GeV to the TeV range. The leptonic model consists of a low-energy Maxwellian distribution for the electrons. The hadronic model parameters are those found by \cite{LiChenW41Had}. The spectra of electrons $e$ and protons $p$ are represented by a power law of slope $\Gamma_{inj}$ with an exponential cutoff at energy E$_{c,e}$ and E$_{c,p}$ respectively.}
                        \label{Tab:ParamWholeSrc}
                        \begin{tabular}{lcc}
                \hline
                \hline
				Fit parameters			&	Leptonic model and Maxwellian		& 	Hadronic model				\\
											&	temperature at 66.9$\pm$0.3 GeV		&	from \cite{LiChenW41Had}	\\
				\hline
			$\tau$ (kyr)					&	10										&	100							\\
				\hline
			B ($\mu$G)					&	$14.1\pm0.8$						&	10								\\
				\hline
			W$_e$ or W$_p$ ($\times10^{49}$ erg)\footnote{Total energy in electrons or protons (depending on the considered model)}
											&	0.24$\pm$0.05							&	10.0							\\
                        \hline
			$\Gamma_{\textrm{inj}}$				& $-2.6\pm0.3$							&	$-2.2$							\\
				\hline
			E$_{c,e}$ or E$_{c,p}$ (TeV)	&	13.4$\pm$3.9					&	50								\\
				\hline
                        \end{tabular}
                \end{center}
                \end{minipage}
                \renewcommand{\footnoterule}{}
                \end{center}
                 \end{table*}
			
			\subsubsection{Only the TeV point-like component is powered by the putative pulsar}
			In an alternative PWN scenario, we assume that only the TeV point-like component (spatially coincident with the X-ray point-like source) is produced by the PWN powered by the putative pulsar.
			Under this assumption, the $\gamma$-ray efficiency is estimated to be $L_{\textrm{TeV}}/\dot E \in [0.1\%, 1\%]$.
			This is consistent with the range of efficiencies (0.01\% -- 10\%) found for several VHE PWNe candidates \citep{HessPwn}.
			
			A simple one-zone leptonic model has been considered and displayed in Figure \ref{Fig:SpectGeVTeVSingle} (dashed line).
			The TeV data have been computed from the normalization of those found in the central extraction region and displayed in Figure \ref{Fig:GeVTeVSpectra}. The calculation of the normalization is detailed in Section \ref{Sec:HessSpec}.
			Despite the large spectral uncertainties, a SED modeling has been performed to estimate the physical parameters involved.
			The model assumes a distribution of accelerated electrons cooling radiatively by means of synchrotron radiation and inverse-Compton scattering, a distance of 4.2 kpc, and an age of $10^4$ yr for the pulsar.
			This age is lower than that estimated for W41, but still in the estimated range for $\tau_c$.
			A good representation of the radio upper limits and X-ray data together with the GeV point-like source upper limits and TeV central component is obtained for an injection spectral index $\Gamma_{inj}=-2.0$, a magnetic field of $B\simeq15~\mu$G. The total energy injected to leptons ($W_e\simeq 3.0\times10^{47}$ erg) represents $\sim10\%$ (or 100\%) of the rotational energy that can be injected by the pulsar over its entire lifetime assuming a constant rotational power of 10$^{37}$ erg~s$^{-1}$ (or 10$^{36}$ erg~s$^{-1}$, respectively), favoring once again a spin-down power higher than 10$^{36}$ erg~s$^{-1}$ for the powering pulsar.
			
			In this scenario, the TeV point-like source is assumed to be unrelated to the TeV extended component.
			Thus, the extended structure seen by \emph{Fermi} LAT and H.E.S.S. may have two possible origins.
			Electrons accelerated in the past and affected by radiative losses could create a relic nebula emitting in \g-rays.
			This ancient PWN would be too faint to be detected in X-rays.
			An energy-dependent morphology of the \g-ray emission from 1.8 GeV up to 100 TeV might suggest the electron cooling in the PWN.
			However, without any significant variation of the intrinsic sizes of the \g-ray source in this energy range or variation of the spectral slope at VHE energy, this scenario tends to be disfavored.
			The other possibility suggests that the emission from the interaction of the SNR with the MC is in the line of sight of the PWN \g-ray emission. But the chance probability of detecting two TeV sources in spatial coincidence in the part of the Galactic plane covered by H.E.S.S. ($140^\circ\times6^\circ$) has been estimated as $p\simeq9.4\times10^{-4}$ from the formula in \cite{ProbCoincid}. Nevertheless, this configuration may happen as in the case of W51C where the TeV source seems to be powered by a PWN and a SNR/MC interaction \citep{W51CMagic,W51CHess}.

		\subsection{SNR-MC interaction}
			Using high-resolution CO observations, \cite{AgeW41} suggested that a MC is located close to the SNR W41 and could even be interacting with the SNR, as proposed by \cite{HESSJ1834_MAGIC}.
			The \g-ray emission detected by H.E.S.S. (as seen in Figure \ref{Fig:W41AllE}) is spatially coincident with the $^{13}$CO high-density region extracted from GRS data \citep{GRSData}, as observed for other interacting SNRs at TeV energy.
			Direct evidence that the shock wave of W41 is interacting with a MC has been discovered recently.
			\cite{OHmaserW41} found two regions with OH maser emission close to the center of the remnant.
			Only the brightest maser at $l,b = (23.26^\circ, -0.31^\circ)$ has been displayed in Figures \ref{Fig:W41AllE}, \ref{Fig:GeVTeVMorpho} and \ref{Fig:Fermi}.

\begin{figure}[t!]
					\centering
					\includegraphics[width=0.5\textwidth]{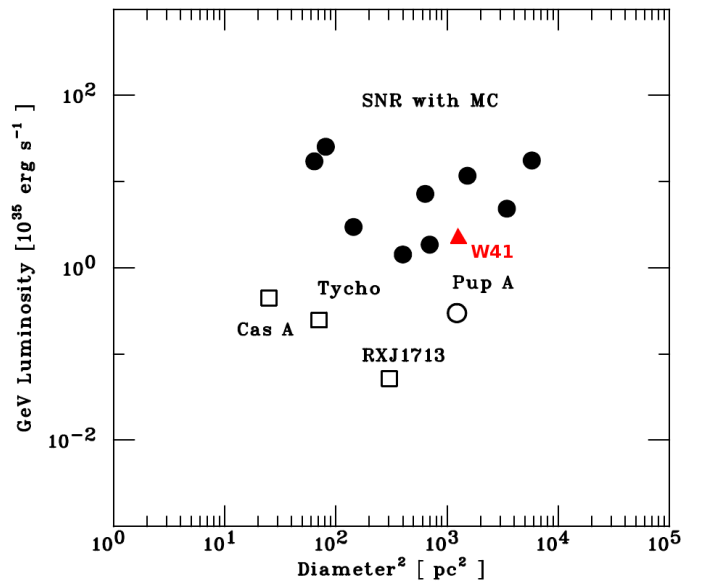}
					\caption{GeV luminosity (0.1-100 GeV) as a function of diameter squared for each of the SNRs detected with \emph{Fermi} LAT \citep{LgamDiamSnr}. Filled circles correspond to the SNRs interacting with molecular clouds. Rectangles represent young SNRs. Puppis A, a middle-aged SNR without clear molecular interaction, is shown as an open circle. The source coincident with the SNR W41 is marked by a red triangle.}
					\label{Fig:LgamDiamSnr}
				\end{figure}

			\emph{Fermi} LAT observations have already led to the detection of \g-ray emission from several SNRs.
			Interestingly, SNRs interacting with MCs constitute the dominant class of \g-ray luminous SNRs, and several evolved interacting remnants (\emph{i.e.} W51C, W44, IC443 and W28) are associated with extended GeV emission \citep{W51C_Fermi, W44_Fermi, IC443_Fermi, W28_Fermi}.
			For W41, the estimated luminosity $L(0.1-100~\textrm{GeV})$ assuming $d=4.2$ kpc is similar to that of other interacting SNRs detected by \emph{Fermi} as shown in Figure \ref{Fig:LgamDiamSnr} \citep{LgamDiamSnr}.
			In this context, \cite{LiChenW41Had} proposed an accumulated diffusion model in which the MC size is taken into account and where the diffusion distance of the protons injected varies as the remnant expands.
			This illuminated cloud model reproduces the GeV and TeV emission using a reasonable energy budget: the total energy injected up to $E_{\textrm{max}}\simeq50$ TeV is $W_p\simeq10^{50}$ erg.
			The set of parameters is summarized in Table \ref{Tab:ParamWholeSrc} and reproduced in Figure \ref{Fig:SpectGeVTeVSingle} (dash-dotted line). This model has been obtained from the spectrum published in \cite{Icrc11W41}.
			The flat spectrum seen by \emph{Fermi} up to several tens of GeV implies a break at $\sim100$ GeV, which is unusual for interacting SNRs.
			This led \cite{LiChenW41Had} to use a cloud as large as the shock wave radius and a diffusion coefficient lower than the average Galactic value.
			This may happen if the interstellar medium is highly ionized by cosmic rays.

\section{Conclusion}
	Follow-up observations have been performed from 1.8 GeV to up to 100 TeV in the direction of the SNR W41 to provide new clues on the origin of the \g-ray emission.
	In the VHE domain, the H.E.S.S. data exhibit a two-component morphology: a central point-like source and an extended component.
	The detailed study of \emph{Fermi} LAT HE \g-ray observations reveals extended emission which has an intrinsic size ($0.15^\circ$) similar to that of the TeV extended component ($0.17^\circ$).
	The GeV and TeV emissions are spatially coincident with a candidate pulsar located at the center of the remnant.
	However, no pulsations in radio, X-rays or HE \g-rays have been detected.
	Different scenarios have been investigated to explain the \g-ray emission.
	
	First, the spatial coincidence of the best-fit position of the \g-ray sources with an X-ray CCO possibly surrounded by a PWN strengthens the PWN scenario. The potential CCO is able to power the whole \g-ray emission only if its current spin-down is close to (or even higher than) $10^{37}$ erg s$^{-1}$. Then, the flat GeV spectrum can only be reproduced assuming a second low-energy electron population. These two aspects are very similar to the case of the PWN candidate HESS~J1640-465 \citep{J1640}.

	Second, the extended GeV and the whole TeV sources are coincident with a high-density region traced by $^{13}$CO. A model based on the assumption of an interaction between W41 and the nearby MC reproduces the spectral data with a reasonable energy budget. We may notice that due to the flat \emph{Fermi} LAT spectrum, an unusually high energy spectral break at $\sim100$ GeV is required. However, the interacting scenario is reinforced by the recent discovery of OH (1720 MHz) maser lines coincident with the remnant and compatible with the associated cloud velocity.
	
	The last possibility would be that the TeV emission has two different sources.
	The point-like component would originate from the PWN powered by the putative pulsar, while the extended TeV component which matches the GeV source would be created by the interaction between the remnant and the cloud.
	In this case, the spatial coincidence between the point-like and the extended VHE \g-ray emission would occur by chance, but the probability is very low.

	Nevertheless, the hadronic scenario seems to explain both GeV and TeV extended emission, as argued in this work. This origin is supported by the recent maser detection and theoretical studies of this \g-ray source.
	
	A clear identification of the extended component in X-rays could help to investigate the hypothesis of a PWN, but given the low-energy flux expected from typical relic PWNe, this appears to be a serious observational challenge.
	Deeper VHE observations with the Cherenkov Telescope Array (CTA) might also allow higher energy resolution studies and a firm identification of the origin of the VHE emission, in particular through morphological analysis.
	Concerning the \emph{Fermi} LAT analysis, the next improvement of the Galactic diffuse emission model and the IRFs might ameliorate the study of the GeV \g-ray emission at lower energies in order to better constrain the modeling of the spectral energy distribution.

\section*{{\small Acknowledgments
{\small
The support of the Namibian authorities and of the University of Namibia in facilitating the construction and operation of H.E.S.S. is gratefully acknowledged, as is the support by the German Ministry for Education and Research (BMBF), the Max Planck Society, the French Ministry for Research, the CNRS-IN2P3 and the Astroparticle Interdisciplinary Programme of the CNRS, the U.K. Science and Technology Facilities Council (STFC), the IPNP of the Charles University, the Polish Ministry of Science and Higher Education, the South African Department of Science and Technology and National Research Foundation, and by the University of Namibia. We appreciate the excellent work of the technical support staff in Berlin, Durham, Hamburg, Heidelberg, Palaiseau, Paris, Saclay, and in Namibia in the construction and operation of the equipment.\\
The \textit{Fermi} LAT Collaboration acknowledges generous ongoing support from a number of agencies and institutes that have supported both the development and the operation of the LAT as well as scientific data analysis. These include the National Aeronautics and Space Administration and the Department of Energy in the United States, the Commissariat \`a l'Energie Atomique and the Centre National de la Recherche Scientifique / Institut National de Physique Nucl\'eaire et de Physique des Particules in France, the Agenzia Spaziale Italiana and the Istituto Nazionale di Fisica Nucleare in Italy, the Ministry of Education, Culture, Sports, Science and Technology (MEXT), High Energy Accelerator Research Organization (KEK) and Japan Aerospace Exploration Agency (JAXA) in Japan, and the K.~A.~Wallenberg Foundation, the Swedish Research Council and the Swedish National Space Board in Sweden.
Additional support for science analysis during the operations phase is gratefully acknowledged from the Istituto Nazionale di Astrofisica in Italy and the Centre National d'\'Etudes Spatiales in France. }}}


\end{document}